\documentclass[12pt]{article}
\input epsf.sty
\usepackage{hyperref}
\usepackage{amssymb}
\usepackage{amsmath}
\usepackage[nosort]{cite}
\usepackage[normalem]{ulem}

\newcommand{\be}{\begin{equation}}
\newcommand{\ee}{\end{equation}}
\newcommand{\ben}{\begin{eqnarray}\displaystyle}
\newcommand{\een}{\end{eqnarray}}

\newcommand{\sectiono}[1]{\section{#1}\setcounter{equation}{0}}

\def\sqr#1#2{{\vcenter{\vbox{\hrule height.#2pt
         \hbox{\vrule width.#2pt height#1pt \kern#1pt
            \vrule width.#2pt}
         \hrule height.#2pt}}}}

\numberwithin{equation}{section}

\begin{document}

\begin{center}
{
\Large{\bf ON THE BLACK HOLE INTERIOR  \\\vspace{4mm}
IN STRING THEORY}}

\vspace{10mm}

\textit{Roy Ben-Israel$\,^1$, Amit Giveon$\,^2$, Nissan Itzhaki$\,^1$ and Lior Liram$\,^1$}
\break

$^1$ Physics Department, Tel-Aviv University, Israel	\\
Ramat-Aviv, 69978, Israel \\

$^2$ Racah Institute of Physics, The Hebrew University \\
Jerusalem, 91904, Israel
\break

\vspace{10mm}

%\vspace{10mm}

\end{center}

\begin{abstract}

The potential  behind the horizon of an eternal  black hole in classical theories is described in terms of data that is available to an external observer -- the reflection coefficient of a wave that  scatters on the black hole.
In GR and perturbative  string theory (in $\alpha'$), the potential is regular at the horizon and it blows up at the singularity. The exact reflection coefficient, that is known for the $SL(2,\mathbb{R})_k/U(1)$ black hole and includes non-perturbative $\alpha'$ effects, seems  however to imply that there is a highly non-trivial structure just behind   the horizon.

\end{abstract}

%\newpage

\baselineskip=18pt

%\tableofcontents

%\newpage
%\renewcommand{\theequation}{\thesection.\arabic{equation}}
%\sectiono{Introduction }
\bigskip
\newpage

\sectiono{Introduction}

Describing the interior of a black hole (BH) and determining the fate of an infalling observer
is a difficult task  in string theory. The observables in string theory are on-shell quantities -- S-matrix elements in flat space-time and correlation functions at the boundary of an AdS background. As such they live at infinity and are natural from the point of view of an external observer that by causality has no access to the BH interior.

The standard way to address this challenge is to find an off-shell effective action that yields the same S-matrix elements as string theory and to look for BH solutions in this effective action. At low energies (relative to the string mass) the relevant effective action is supergravity (SUGRA) which respects the equivalence principle. Consequently, the horizon of a BH in SUGRA is smooth.   At higher energies corrections to SUGRA become relevant. Perturbative corrections (both in $\alpha'=l_s^2$ and in the string coupling, $g_s$) can be described in terms of higher orders terms (roughly of the type $R^2$) \cite{Gross:1986iv} that also respect the equivalence principle and are small at the horizon of a large BH. Therefore, they can have an important effect only at the singularity (see e.g. \cite{Callan:1988hs}). In particular, the horizon remains smooth when perturbative corrections are taken into account.

What happens in the full string theory, including all corrections, perturbative or otherwise? In particular, can non-perturbative stringy effects violate the equivalence principle and render the horizon singular?
One might view this as a philosophical question since we cannot solve string theory exactly. We do not have an effective action that takes account of all non-perturbative corrections that can be used to test the equivalence principle.

There is, however, a setup in which
this question becomes quite precise. Some correlation functions were calculated exactly in the coset conformal field theory (CFT)
$SL(2,\mathbb{R})_k/U(1)$ \cite{Teschner:1999ug}. These  include the exact reflection coefficient associated with $k$ near-extremal NS5-branes in the classical limit, $g_s=0$. The goal of this note is  to take advantage of this exact stringy data in order to determine whether the horizon is regular or not in this setup. More precisely, we wish to calculate the potential outside and behind the horizon in terms of the reflection coefficient.

It is natural to expect  that for large $k$ the answer must be that the horizon is smooth and that
large deviations from the SUGRA potential take place only near the singularity. This expectation is based on the fact that  we consider a classical setup while arguments  so far for non-trivial structure at the BH horizon are quantum mechanical
\cite{Itzhaki:1996jt,Braunstein:2009my,Mathur:2009hf,Almheiri:2012rt,Marolf:2013dba,Polchinski:2015cea}.
However, as we shall see, things are  more interesting.

The paper is organized as follows: in the next section we describe how to calculate the potential behind the horizon of an Eternal Black Hole (EBH) in terms of data that is available to an external observer -- the reflection coefficient. In section 3 we show  that in SUGRA (including perturbative corrections) the potential  blows up only at the singularity while in the exact
$SL(2,\mathbb{R})_k/U(1)$ BH model the potential blows up just behind the horizon. In section 4 we discuss some aspects associated with coarse-graining the potential behind the horizon. We summarize in section 5.

\sectiono{The potential behind the horizon}

In this section we show how to determine the potential behind the horizon of an EBH in terms of data that is available to an external observer -- the reflection coefficient associated with a scattering of a wave, with some momentum $p$, on a classical black hole, $R(p)$.

Let us suppose that we know $R(p)$.
At least in classical cases, which we focus on here, we expect $R(p)$ to drop exponentially fast with $p$ since most of the wave gets absorbed by the black hole. Still the reflected wave contains information about the surrounding of the black hole. In particular, one can ask what kind of  a potential, $V(x)$,  in a Schr{\"o}dinger-like setup,
\be\label{sch}
-\partial_x^2 \psi(x)+V(x) \psi(x)=p^2 \psi(x),
\ee
yields $R(p)$. In quantum mechanics this is known as the inverse scattering problem.

In GR  (\ref{sch}) can be found by transforming the Klein-Gordon equation in the EBH background into a Schr{\"o}dinger equation. In this case
$x$ is the tortoise coordinate, where $x=\infty$ is the asymptotic region and $x=-\infty$ is the horizon. From the  point of view of GR, the input is the BH background and the output is the reflection coefficient.

In string theory the logic is reversed: the basic quantities that we know how to calculate are on-shell scattering amplitudes, from which we can read $R(p)$. Once  known, one can ask what kind of a background yields $R(p)$. In particular, we wish to know if this background admits a smooth  horizon. To do so we solve the inverse scattering problem and find the potential $V(x)$. The background is singular when $V(x)$ blows up. In other words we can view $V(x)$ as a singularity detector.
A refinement of this statement appears at the end of this section.

The range of $x$  indicates that the only information about the EBH we can extract {\it directly} this way is about the region outside the black hole. This is as it should be since causality implies that the reflection should take place outside the horizon for the reflected wave to make it back to infinity. To find the potential behind the horizon, which we denote by $V_{int}(x)$, we can  analytically continue $V(x)$.
From $V_{int}(x)$ we can tell if there is a singularity behind the eternal BH horizon and where it is  located.  In GR  $V_{int}(x)$ can be found  by writing the Klein-Gordon equation behind the horizon in a Schr{\"o}dinger equation form. As will be discussed momentarily this equation can also be found from the  analytic continuation of the Schr{\"o}dinger equation  outside the horizon.

Our goal here is to  find the relation between $R(p)$ and $V_{int}(x)$.  Such a relation  can be used in order to determine in the exact $SL(2,\mathbb{R})_k/U(1)$ BH model the potential behind the horizon. In particular, it can be used to tell if something unusual happens to $V$ when we cross the EBH horizon. To do so we
first recall how the potential outside the BH and the reflection coefficient are related.

In 1D, when there are no bound states, and when the potential decays sufficiently fast as $x \rightarrow \pm \infty$, the potential is related to the scattering data in the following way  (see e.g. \cite{Deift:1979}),
\be\label{q}
	V(x)=\frac{2i}{\pi}\int_{-\infty}^{\infty} p \, R(p)\,\frac{T^*(p)}{T(p)}(\psi^*(p,x))^2 dp.
\ee
$T(p)$ is the transmission coefficient and $\psi(p,x)$ is the Jost solution to the Schr{\"o}dinger equation that asymptotically behaves as
\be\label{jost}
  \psi(p,x) \sim
  \begin{cases}
	   e^{-ipx}, & x\rightarrow -\infty,\\
	   \frac{1}{T(p)}e^{-ipx}+\frac{R(p)}{T(p)}e^{ipx}, & x\rightarrow \infty.
  \end{cases}
\ee
In the case of EBHs $V(x)$ is smooth and its maximal value is of the order of the curvature. Consequently, for wavelengths much smaller than the curvature scale, $T(p)$  approaches $1$ and $R(p)$ decays exponentially fast. In fact, both for  Schwarzchild EBH (see appendix A) and for $SL(2,\mathbb{R})_k/U(1)$ EBH (see next section) the asymptotic behaviour of the reflection coefficient is
\be\label{l}
   R(p)\sim e^{-\frac{\beta}{2} |p|},
\ee
where $\beta$ is the inverse temperature.

The potential at the horizon (from the outside) is smooth.
Combining (\ref{q}) and (\ref{jost}) we see that near the horizon it takes the form
\be\label{horizonout}
V_{Hor-out}=\lim_{x\to-\infty}\frac{2i}{\pi}\int_{-\infty}^{\infty} p \, R(p)\,\frac{T^*(p)}{T(p)} e^{2 i p x} dp.
\ee
We already know that $V_{Hor-out}$ is smooth. Still it is worthwhile to see how this is consistent with (\ref{horizonout}). A divergence in (\ref{horizonout}) can come only from the UV. However, since at the UV $T(p)\to 1$, we get from (\ref{l}) that
\be\label{horoutUV}
V_{Hor-out}^{UV}\sim \lim_{x\to-\infty} \int p \,e^{-\frac{\beta}{2} |p|} e^{2 i p x} dp,
\ee
which converges rapidly.

What happens when we cross the horizon? The relation between the tortoise and Kruskal coordinates
\be\label{tortoiseKruskalRelation}
	x=\frac{\beta}{4\pi} \log(uv),
\ee
implies that  crossing the horizon amounts to taking
\be\label{b}
	x\to x \pm \frac{i\beta}{4}.
\ee
The `$\pm $' is due to the two simplest possible branch choices in (\ref{tortoiseKruskalRelation}).

In the interior  the horizon is at $\text{Re}\,(x)=-\infty$ and the singularity at $\text{Re}\,(x)=0$.\footnote{The asymptotic region behind the singularity, that plays no role here, is at $\text{Re}\,(x)=\infty$.} Eq. (\ref{b}) implies that the potential behind the horizon, $V_{int}(x)$, is related to $V(x)$ via
\be\label{ll}
V_{int}(x)=V(x \pm i\beta/4).
\ee
Hence the potential just behind the horizon reads
\be\label{horizonin}
V_{Hor-In}=\lim_{x\to-\infty}\frac{2i}{\pi}\int_{-\infty}^{\infty} p \, R(p)\,\frac{T^*(p)}{T(p)} e^{2 i p x} \cosh(\beta p/2) dp.
\ee
We emphasise that $T(p)$ and $R(p)$ in this equation are the transmission and reflection coefficients associated with scattering {\it outside} the EBH. The $\cosh(\beta p/2)$ is the only difference between the potential inside and the potential outside. This follows from (\ref{ll}). Note that we keep the two branches of (\ref{ll}) which together with
\be\label{kob}
R^{*}(p)=R(-p),~~~~T^{*}(p)=T(-p)
\ee
make the reality of $V_{Hor-In}$ manifest.

It is clear that $V_{Hor-out}$ is finite. Again a divergence can appear at the UV where $T(p)\to 1$. If (\ref{l}) holds we get that
\be\label{horinUV}
V_{Hor-In}^{UV}\sim \lim_{x\to-\infty} \int p \,e^{-\frac{\beta}{2} |p|} e^{2 i p x} \cosh(\beta p/2) dp,
\ee
that is smooth at the horizon.

There are a couple of comments in order at this stage:\\
$\bullet$ Both $V_{Hor-In}$ and  $V_{Hor-Out}$ vanish when $x\to-\infty$. Hence there is no discontinuity as we cross the horizon. In the next section we show this explicitly for the $SL(2,\mathbb{R})_k/U(1)$ EBH. This is a general result  that follows from the equivalence principle held in GR and perturbative string theory.\\
$\bullet$ If we relax the condition  $x\to-\infty$ in (\ref{horinUV}) we see that its RHS blows up at $x=0$ exactly where the singularity is located. This is not a coincidence. At the UV the wave function $\psi(x,p)$ is well approximated by $e^{ipx}$ which implies that the contribution of the UV modes to the potential is given by the RHS of (\ref{horinUV}) for any $x$. As is evident from the RHS of (\ref{horinUV}) the UV modes are sufficient to detect the singularity.

In the next section we shall see that in the exact $SL(2,\mathbb{R})_k/U(1)$ EBH (\ref{l}) does not hold and that consequently there is a non-trivial structure just behind the horizon. Before we do so we make a slight refinement.\\
{\bf A Refinement}

Below we refine  the definition of the singularity detector. This refinement does not affect any of the conclusions that follow.

To motivate the refinement we consider a free massive mode in Rindler space.
Rindler space is merely a reparametrization of Minkowski space, and so an ideal singularity detector should vanish everywhere in Rindler space.  However, as we now show,  $V(x)$ does not vanish in Rindler space-time. In fact, it blows up at infinity. To see this we consider the massive Klein-Gordon equation of a mode with energy $\omega$ in the Rindler metric
\be
   ds^2=-\left(\frac{2\pi}{\beta}\right)^2 \rho^2 dt^2 + d\rho^2,
\ee
where the Unruh (inverse) temperature, $\beta$, is written explicitly. It  takes the form
\be
\frac{1}{\rho}\left[ \frac{\beta^2}{4\pi^2 \rho} \omega^2 \phi+\partial_{\rho} (\rho\partial_{\rho}\phi) \right]=m^2\phi.
\ee
To put it in a Schr{\"o}dinger form, we switch to the tortoise coordinate
\be
   x=\frac{\beta}{2\pi} \log \left(\frac{2\pi \rho}{\beta}\right),
\ee
to find
\be
-\partial_x^2 \phi(x)+e^{\frac{4\pi}{\beta} x} m^2 \phi(x)=\omega^2\phi(x).
\ee
Therefore
$
V(x)=e^{\frac{4\pi}{\beta} x} m^2,
$
which blows up at infinity.  Clearly, the fact that the potential blows up at infinity does not indicate there is a singularity there. To fix this we define
\be \label{dramaParameter}
  D(x)= \partial_x (e^{-\frac{4\pi}{\beta} x} V(x)).
\ee
The factor $e^{-\frac{4\pi}{\beta} x}$ is basically the redshift  factor between the Rindler and Minkowski observers.

Indeed in Rindler space $D(x)$ vanishes and in the BH background, much like $V(x)$, it blows up only at the singularity. 
% A related reason to prefer $D(x)$ over $V(x)$ is that it captures better the tidal forces an infalling observer experiences.
In the following section we show that $D(x)$ (and also $V(x)$) blows up at the horizon of the $SL(2,\mathbb{R})_k/U(1)$ BH. One may wonder if some alternative definition of $D(x)$ would not be divergent at the horizon. While this is true, the definition \eqref{dramaParameter} could be related to tidal forces acting on infalling observers which is a measurable physical quantity. As such, it is a valid singularity detector.

\section{The $SL(2,\mathbb{R})_k/U(1)$ BH}

In this section we consider classical ($g_s=0$) EBH associated with the $SL(2,\mathbb{R})_k/U(1)$ model. First, we ignore non-perturbative stringy corrections to the reflection coefficient and obtain the expected result that the horizon is smooth. Then we show that the full stringy  reflection coefficient, that includes non-perturbative effects, alters this conclusion quite dramatically.

\subsection{Semi-classical description}

At the semi-classical level the $SL(2,\mathbb{R})_k/U(1)$ EBH can be described by the following background \cite{Elitzur:1991cb,Mandal:1991tz,Witten:1991yr}
\be\label{background}
ds^2=-\tanh^2 \left(\frac{\rho}{\sqrt{2 k}}\right) dt^2+d\rho^2 ,~~~~~
\exp(2 \Phi )= g_{0}^2\frac{1}{\cosh^2 \left(\frac{\rho}{\sqrt{2 k}}\right)},
\ee
where $\Phi$ is the dilaton.  We are interested in the classical limit of the theory, so we take $g_0\to 0$.

As mentioned above, once we have a solution associated with an EBH we can verify directly that the horizon is smooth.
In terms of $(t,\rho)$ this background is defined only for the exterior region with the horizon at $\rho = 0$, but it can be extended to the region behind the horizon, as usual, by switching to Kruskal coordinates,
\be\label{uv}
u=\sinh\left(\frac{\rho}{\sqrt{2k}}\right) e^{t / \sqrt{2k}},~~~v=\sinh\left(\frac{\rho}{\sqrt{2k}}\right) e^{-t / \sqrt{2k}},
\ee
in terms of which the background takes the form
\be
ds^2=\frac{2k}{1+uv}dudv,~~~\Phi-\Phi_0=-\frac{1}{2}\log(1+uv).
\ee
We see that the background is smooth at the horizon ($uv=0$) and that it is singular at $uv=-1$.
This, of course, is well known. Here we wish to show that $V(x)$ (or $D(x)$) lead to the same conclusion.

The starting point is the reflection coefficient from which we can deduce the potential. At the perturbative level, the reflection coefficient can be determined from the underlying $SL(2,\mathbb{R})$ structure of the $L_0$ and $\bar{L}_0$ generators \cite{Dijkgraaf:1991ba}. In the supersymmetric case it reads
\be\label{R_gravity}
R_{SUGRA}(p)=
\frac{\Gamma(i\sqrt{2k}p)\Gamma^2\left( \frac12(1-i\sqrt{2k}p-i\sqrt{2k}\omega)\right) }
     {\Gamma(-i\sqrt{2k}p)\Gamma^2\left( \frac12(1+i\sqrt{2k}p-i\sqrt{2k}\omega )\right) },
\ee
where $p$ is the momentum in the $\rho$ direction as measured at infinity and  $\omega$ is the energy (associated with $t$) of the wave.
They are related by the on-shell condition
\begin{equation}\label{OnShellCondition}
 \omega = \pm\sqrt{p^2+\frac{1}{2k}}.
\end{equation}
The branch choice for the square root is such that for $\omega,p \in \mathbb{R},~\text{sign}(\omega)=\text{sign}(p)$.
This assures that the reflection coefficient, as required, has the property (\ref{kob}) that is used below.

At large $p$ (or $\omega$) we get (\ref{l}) (with $\beta=2\pi \sqrt{2k}$). Therefore the discussion in the previous section  guarantees that the potential blows up only at the singularity and that, in particular,  the horizon is regular. Still it is  worthwhile to do this exercise here  since, as we shall see, non-perturbative corrections will alter this conclusion quite severely. Note that the exact same expression for the reflection coefficient is obtained by solving the Klein-Gordon equation in the background (\ref{background}). This implies that there are no perturbative $\alpha'$ corrections in the supersymmetric case \cite{Bars:1992sr,Tseytlin:1993my}.

The potential that gives this reflection coefficient is
\be \label{VofX}
	V(x)=\frac{1}{2k}\left( 1-\frac{1}{\left(1+e^{\sqrt\frac2k x}\right)^2}\right).
\ee
Much like in Schwarzschild BH, the potential goes to zero exponentially fast at the horizon $x\to-\infty$. However, unlike in Schwarzschild BH it does not go to zero at infinity, but rather to a positive constant. This is a feature that is related to the linear dilaton at infinity and is not expected to affect the physics at the horizon. To find its behaviour behind the horizon
we write down the tortoise coordinate  in terms of the Kruskal coordinates \eqref{uv},
\begin{equation}
  x = \sqrt{\frac{k}{2}}\log (uv),	
\end{equation}
which gives
\begin{equation}
V(u,v) = \frac1{2k} \left(1-\frac1{(1+uv)^2}\right).
\end{equation}
We see that except at the singularity, where it diverges, the potential is smooth everywhere. In particular, it is continuous across the horizon where it vanishes.

$D(x)$  \eqref{dramaParameter} associated with this potential is,
\begin{equation}
   D(x)= \frac{\tanh \left(\frac{x}{\sqrt{2k}}\right)-2}{4 \sqrt{2} k^{3/2}\cosh^2\left(\frac{x}{\sqrt{2k}}\right)}.
\end{equation}
It is clear that everywhere outside the BH $D(x)$ is small. As for the potential, we write it in terms the Kruskal coordinates to see how it behaves behind the horizon, 
%We see that $D(x)$ is small outside the BH. To find its behaviour behind the horizon
%we write down the tortoise coordinate  in terms of the Kruskal coordinates \eqref{uv},
%\begin{equation}
%  x = \sqrt{\frac{k}{2}}\log (uv),	
%\end{equation}
%which gives
\begin{equation}
  D(u,v)= -\frac{uv (3+ uv)}{\sqrt{2} k^{3/2}(1+ uv)^3}.
\end{equation}
Hence for large $k$ we see that $D(x)$ is small everywhere but at the singularity.

\subsection{Perturbative $\alpha'$ corrections}

The $SL(2,\mathbb{R})_k/U(1)$ model also illustrates neatly that perturbative  $\alpha'$ corrections are negligible at the horizon of a large BH. In particular, they do not render the horizon singular.

In the bosonic case the reflection coefficient obtained via the underlying $SL(2)$ structure of the $L_0$ and $\bar{L}_0$ generators differs a bit from (\ref{R_gravity}) \cite{Dijkgraaf:1991ba}.
This difference, that is due to the shift $k\to k-2$, implies that the classical background (\ref{background}) receives perturbative $\alpha'$ corrections. To determine the modified background, one seeks a background in which the Klein-Gordon equation leads to the corrected  reflection coefficient \cite{Dijkgraaf:1991ba}.
We do not present the details of this calculation (that can be found in \cite{Dijkgraaf:1991ba}) since, not too surprisingly,
this background is regular at the horizon.
In the next subsection we shall see that non-perturbative $\alpha'$ corrections are more interesting.

\subsection{Exact description}

A nice feature of the $SL(2,\mathbb{R})_k/U(1)$ EBH is that, much like in the Liouville model \cite{Zamolodchikov:1995aa}, the reflection coefficient can be calculated exactly on the sphere \cite{Teschner:1999ug}. It reads
\be\label{ex}
  R_{exact}(p) =  R_{SUGRA}(p) R_{non-per}(p),\ee
with
\be
  R_{non-per}(p) = -\frac{\Gamma\left( i\sqrt{\frac2k}p\right) }
                          {\Gamma\left(-i\sqrt{\frac2k}p\right)}, \ee
where $R_{SUGRA}$ is given in \eqref{R_gravity}.

The exact reflection coefficient in the $SL(2,\mathbb{R})_k/U(1)$ EBH is determined by the exact reflection coefficient in the Euclidean $H_3^{+}$ \cite{Teschner:1999ug}. There is no transmission coefficient in  $H_3^{+}$. Therefore we do not know the exact transmission coefficient in the $SL(2,\mathbb{R})_k/U(1)$ EBH. It is natural to assume that for $p^2 \gg 1/k$
\be\label{T}
T(p)\to 1,
\ee
since this is a general result that follows from having a smooth potential outside the EBH that is bounded by the curvature that scales like $1/k$. It is possible that in the exact string theory (\ref{T}) does not hold, but that would mean that non-perturbative effects in classical string theory affect drastically the region outside the EBH. We find this hard to believe and so we assume (\ref{T}).

We would like to argue now  that  (\ref{ex}) and (\ref{T}) imply that there is a singularity just behind the horizon. This too is a bit hard to believe, but it is less dramatic  than the alternative that follows from a violation of (\ref{T}).

At first sight it is hard to see how such a startling result can come about. The non-perturbative correction, $R_{non-per}$, is merely a phase. Instead of (\ref{l}), the leading UV behaviour is
\be
   R(p)\sim 2e^{-\frac{\beta}{2} |p|+i\theta(p)}.
\ee
It is clear from (\ref{horoutUV}) that $V_{Hor-Out}$ still converges rapidly (which is part of our assumption in reaching (\ref{T})).
On the other side of the horizon  the UV contribution
to $V_{Hor-In}$ reads
\be\label{vintdd}
V^{UV}_{Hor-In}(x)= \lim_{x\to-\infty}\frac{2i}{\pi}\int p \, e^{i (2 p x +\theta(p))} dp.
\ee
If $\theta(p)$ was a function that approaches a constant at large $p$ then the RHS of (\ref{vintdd}) still would have blown up only at the location of the GR singularity ($x=0$) and the effect of $\theta(p)$ was to smear it a bit.
Similarly if at the UV we had $\theta(p)\sim c \, p$ then the singularity in the RHS of (\ref{vintdd}) would have
been shifted by $c$. A combination of these two options, shifting a bit and smearing the singularity, is something
that is natural to expect from $\alpha'$ corrections.

This, however, is not what happens in the exact string theory. Using Stirling's approximation, one finds that for
 $p^2\gg k$ we have
\be\label{stringyPhase}
	-\frac{\Gamma\left( i\sqrt{\frac2k}p\right)}
	     {\Gamma\left(-i\sqrt{\frac2k}p\right)}
	     \sim  i\, \text{sign}(p)\, e^{i\theta(p)},~~~\mbox{with}~~~	
 	\theta(p) \sim  \sqrt{\frac{8}{k}} \, p \log \left(\sqrt{\frac2k} \frac{|p|}{e}\right).
\ee
This means that the rate by which the phase is growing, keeps on increasing indefinitely.
This fact leads to some interesting effects already in the corresponding cigar geometry \cite{Giveon:2015cma,Ben-Israel:2015mda} and consequently  on
 the relevant Hartle-Hawking wave function \cite{Ben-Israel:2015etg}.

Here this implies that the singularity gets expelled all the way to the horizon. This is because the shift of the singularity is dependent on $p$ and it keeps on growing indefinitely at the UV, $c(p) \sim \sqrt{8/k} \log (|p|)$. A more precise way to see this
is to consider the potential behind the horizon, (\ref{horinUV}), that now reads
\begin{equation}\label{Vinside}
V^{UV}_{Hor-in}(x)= \lim_{x\to-\infty}\frac{2i}{\pi}\int_{p^2>k}  p \, e^{i\left( \sqrt{\frac{8}{k}} \, p \log \left(\sqrt{\frac2k} \frac{|p|}{e}\right)+2px\right)} dp,
\end{equation}
%\begin{equation}\label{Vinside}
%V^{UV}_{Hor-in}(x)= \lim_{x\to-\infty}\frac{4}{\pi}\int_{p^2>k}\text{sign}(p)  p \, e^{-i\left( \sqrt{\frac{8}{k}} \, p \log \left(\sqrt{\frac2k} \frac{|p|}{e}\right)+2px\right)} dp,
%\end{equation}
where we made use of \eqref{stringyPhase}.
The integral is controlled by saddle points at
\be\label{sp}
p=\pm \sqrt{\frac{k}{2}}e^{-\sqrt{\frac{k}{2}} x}
\ee
 that give
\begin{equation}\label{VInsindeNearHorizon}
V^{UV}_{Hor-in} \sim k \, e^{-\frac32\sqrt{\frac{k}{2}}x } \cos\left( 2  \, e^{-\sqrt{\frac{k}{2}}x}\right).
\end{equation}
%\begin{equation}\label{VInsindeNearHorizon2}
%V^{UV}_{Hor-in} \sim k \, e^{-\frac32\sqrt{\frac{k}{2}}x } \sin\left( 2  \, e^{-\sqrt{\frac{k}{2}}x}\right).
%\end{equation}
We see that while in perturbative string theory the potential behind the horizon is small (of order $1/k$) until close to the singularity, in the full classical string theory it is large (of order $k$) and it blows up exponentially fast while oscillating with an exponentially large frequency  as we get closer to the horizon.

The same holds for $D(x)$ that reads
\begin{equation}\label{DInsindeNearHorizon}
   D_{Hor-in}(x) \sim k^{\frac32} \, e^{-\frac52\sqrt{\frac{k}{2}}x}
         \sin\left( 2  \, e^{-\sqrt{\frac{k}{2}}x}\right).
\end{equation}
%\begin{equation}\label{DInsindeNearHorizon2}
%   D_{Hor-in}(x) \sim k^{\frac32} \, e^{-\frac52\sqrt{\frac{k}{2}}x}
%         \cos\left( 2 \, e^{-\sqrt{\frac{k}{2}}x}\right).
%\end{equation}
We conclude that while the non-perturbative stringy corrections  in $\alpha'$ have a tiny effect outside the horizon, they render the region just
 behind the horizon singular.

\section{Coarse-graining}

In this section we discuss two aspects associated with coarse-graining the results of the previous section. One from the point of view of an external  observer and the other from the point of view of an infalling observer.

\subsection{An external observer}

Let us suppose that an external observer attempts to check {\it experimentally} that there is, as we claimed, a non-trivial structure just behind the horizon of a classical BH. Assuming there is no flaw in our reasoning, such an observer should measure the reflection coefficient and see if  (\ref{ex}) is correct.
The non trivial effects we discussed come from the deep UV and so a natural  question to ask  is: what could an external observer, that has access to energies below some cutoff $\Lambda$,  concludes experimentally about the structure behind the horizon (without extrapolating her findings to arbitrarily high energies)?

From (\ref{sp}) we see that if we  cut $p$ (or $\omega$) off at $\Lambda$, then the singularity is not pushed all the way to the horizon, but to
\be\label{ao}
x= -\sqrt{\frac2k} \log\left(\frac{\Lambda}{\sqrt{k/2}}\right).
\ee
To understand the physical meaning of this it is useful to switch to $\rho$, the invariant distance from the horizon.
Using (\ref{tortoiseKruskalRelation}) and the fact that beyond the horizon $\rho\to i \rho$ we have
\be\label{oa}
x=\sqrt{2k}\log\left( \sin\left(\frac{\rho}{\sqrt{2k}}\right)\right).
\ee
The classical singularity is at $\rho_{sin}=\pi\sqrt{k/2}$.

Combining (\ref{ao}) and (\ref{oa}) we see that the singularity is at
\be
\log\left(\sin\left(\frac{\rho}{\sqrt{2k}}\right)\right)=-\frac1k \log\left(\frac{\Lambda}{\sqrt{k/2}}\right).
\ee
In the large $k$ limit and for $\Lambda$ that is not exponentially large  the singularity is pushed only slightly away from $x=0$: writing $\rho=\rho_{sin}-\delta\rho$ and expanding in $\frac{\delta\rho}{\sqrt{k}}$ we find
\be\label{ty}
\delta\rho^2=4 \log\left(\frac{\Lambda}{\sqrt{k/2}}\right).
\ee
As long as $\Lambda$ is not exponentially large in some power of $k$, the singularity appears to inflate by a  stringy distance with a mild logarithmic dependence on the cutoff.

This  is reminiscent of the  root mean square
variation of the transverse directions in string theory \cite{Karliner:1988hd}
\be\label{mi}
\langle (\Delta X)^2 \rangle = \alpha' \sum_{n>0} \frac1n
\ee
where the sum is over the stringy modes. This sum diverges, but a finite resolution of the measuring device gives
\be\label{mil}
\langle (\Delta X)^2 \rangle = 2 \log(n_{max}).
\ee
$n_{max}$ plays the role of the cutoff and the factor of $2$ is due to the fact that we work with $\alpha'=2$.  Since the mass of the string at level $n$ is  $M^2=n \, \alpha' $ it is natural to relate $n_{max}$  to $\Lambda^2$ and so
there is an agreement between (\ref{mil}) and (\ref{ty}). It would be nice if this technical agreement could shed light on the origin  of our results; perhaps in relation with  \cite{Susskind:1993aa,Susskind:1993ws,Dodelson:2015toa} in which it was speculated that  (\ref{mil}) might play an important role in BH physics.

To conclude experimentally that the potential is large  at macroscopic distances away from the classical singularity, the external observer should probe the BH with exponentially large energies. Since we work with $g_s=0$, these energies are still negligible compared to the mass of the BH.
This suggests that small but finite $g_s$  could completely modify the picture.  A finite $g_s$ does not induce a cutoff, but it does modify dramatically the physics at energies of the order of $1/g_s$ \cite{Shenker:1995xq}. We are, unfortunately, in no position to comment on that.

\subsection{An infalling observer}

We have just concluded that an external observer will have to reach exponentially large energies in order to conclude that there is structure just behind the horizon. Does this mean that an infalling observer with a smooth wave function is not sensitive to this structure and can fall freely through it?

The fact that  $V(x)$ oscillates wildly seems to support this possibility.
If, for example, we average  $V(x)$ with a Gaussian wave function with some width $\Delta$
\be
V(x, \Delta) = \frac{1}{\Delta \sqrt{\pi/2}} \int dx' \, V(x') \, e^{-\frac{(x-x')^2}{\Delta^2}},
\ee
then it is easy to see that despite the fact that the amplitude of $V$ grows faster than its frequency, $V(x, \Delta)$ does not blow up at the horizon.

However, $x$ is not the coordinate associated with an infalling observer. Near the horizon, the infalling observer coordinates are
\be
U=\sqrt{2k}u,~~~~V=\sqrt{2k}v,
\ee
so that at the horizon we have $ds^2=dU dV$.
Moreover, instead of $V(x)$ we should consider $D(x)$ that is more closely related to the tidal forces an infalling observer would experience. In terms of these coordinates we have
\be
D(U,V) \sim k^{3/2} \left(\frac{U V}{2k}\right)^{-\frac{5 k}{4}} \sin\left(2\left(\frac{U V}{2k}\right)^{-\frac{ k}{2}}\right).
\ee
%\be
%D(U,V) \sim k^{3/2} \left(\frac{U V}{2k}\right)^{-\frac{5 k}{4}} \cos\left(2\left(\frac{U V}{2k}\right)^{-\frac{ k}{2}}\right).
%\ee

Since $D(U,V)$ blows up at finite values of $U$ and $V$ (at $V,U=0$) smearing it with a wave function that is natural for an infalling observer does not wash away the singularity at the horizon. We conclude that the fact there is a non-trivial structure just behind the horizon is something  an infalling observer experiences.

\section{ Summary}

In this note we showed how data that is
available to an external observer -- the reflection coefficient -- can be used  to  calculate the potential behind the horizon of an EBH.
As expected  in perturbative string theory, the potential is small and smooth at the horizon and it blows up only at the singularity.
However, the exact reflection coefficient that is known for the $SL(2,\mathbb{R})_k/U(1)$ EBH appears to suggest that the region just behind the horizon is singular.

At first sight this conclusion appears to be too dramatic   since we considered classical string theory ($g_s=0$) while previous arguments for structure at the horizon are quantum mechanical \cite{Itzhaki:1996jt,Braunstein:2009my,Mathur:2009hf,Almheiri:2012rt,Marolf:2013dba,Polchinski:2015cea}.
However, \cite{Itzhaki:1996jt,Braunstein:2009my,Mathur:2009hf,Almheiri:2012rt,Marolf:2013dba,Polchinski:2015cea}
merely state what should happen for the information to be emitted in the radiation:  the Hawking particles must be on-shell very close to the horizon \cite{Itzhaki:1996jt,Polchinski:2015cea} and not just at infinity. Or alternatively, the Hawking particle and its partner cannot form a pure state \cite{Braunstein:2009my,Mathur:2009hf,Almheiri:2012rt,Marolf:2013dba}. None of these papers, however, explain how this comes about. In other words, they do not find a mistake in Hawking's derivation \cite{Hawking:1974sw} that the information is lost. Rather they argue  which of his conclusions must be wrong for the information to be recovered. It is possible, we believe, that the results presented here could fill in this gap. Namely, the classical non-perturbative stringy effects are the seeds for the quantum effects discussed in \cite{Itzhaki:1996jt,Braunstein:2009my,Mathur:2009hf,Almheiri:2012rt,Marolf:2013dba,Polchinski:2015cea}.

For this to happen we should be able to answer the following question: what is the  origin, in {\em classical} string theory, of the  structure just behind the horizon? Simply put, how come that in classical string theory the EBH horizon is not smooth? Currently we do not have an answer to this
 question. However, we would like to point out that
a related effect occurs in the Euclidean version of the BH --
the cigar geometry \cite{Giveon:2015cma,Ben-Israel:2015mda}. There it is believed that the source is the condensation of the winding tachyon \cite{Kutasov:2005rr,Giveon:2012kp,Giveon:2013ica}.
It is not clear to us what is the Lorentzian analogue of the winding tachyon. This, we think, is likely to be a key ingredient for improving the understanding of the results presented here.

We wish to end by spelling out the various assumptions made in reaching the conclusion that the horizon is singular:

$\bullet$ We assumed that the exact reflection coefficient is given by \eqref{ex}. The original calculation of Teschner was done
for $H_3^{+}$ \cite{Teschner:1999ug} and  simple manipulations (gauging and Wick rotation) give \eqref{ex}.
The results of \cite{Teschner:1999ug} (before and after gauging) were rederived using other methods
(see e.g. \cite{Giveon:1999px,Giribet:2000fy,Giribet:2001ft}).

$\bullet$ We assumed that $T(p)\to 1$ at the deep UV. Violating this would  mean that already outside the horizon of $SL(2,\mathbb{R})_k/U(1)$ EBH non-perturbative effects in classical string theory have dramatic effects.

$\bullet$ We assumed a Schr{\"o}dinger-like setup or equivalently a  Klein-Gordon equation in some background and studied if this
background is regular at the horizon. The surprising results we encountered come from the UV and it is natural to wonder if the
Klein-Gordon equation is the right equation of motion to use. In particular, the effects are due to scales such that
$p^2 \gg k$ (see \eqref{stringyPhase}) and at such high scales other terms might be important. The fact that in perturbative string theory a Klein-Gordon
 equation in a regular background gives the correct reflection coefficient \cite{Dijkgraaf:1991ba} does not
appear to support this possibility. Neither does the Euclidean setup \cite{Ben-Israel:2015mda}.
Nevertheless, it is clearly worthwhile to explore the possibility that corrections to the Klein-Gordon equation at the UV can render the horizon smooth.

$\bullet$ A related assumption is that the relation between the tortoise  and Kruskal coordinates is not modified at the deep UV.
This, we believe, is the same as assuming that the equivalence principle holds.

\section*{Acknowledgments}

We thank D. Kutasov for discussions.
This work  is supported in part by the I-CORE Program of the Planning and Budgeting Committee and the Israel Science Foundation (Center No. 1937/12), and by a center of excellence supported by the Israel Science Foundation (grant number 1989/14). LL is thankful for the support from the Alexander Zaks fellowship.

\begin{appendix}
\section {Schwarzschild BH Reflection Coefficient}

In this appendix we compute the high energy behavior of the reflection coefficient in the case of a Schwarzschild BH and show that it obeys \eqref{l}. As mentioned above, when put in terms of the tortoise coordinate, the Klein-Gordon equation can be reduced to a Schr\"{o}dinger-like equation. For the Schwarzschild metric, the tortoise coordinate is
\begin{equation}\label{SchwarzTortoise}
    r_*=r+2M\log\left(\frac{r}{2M}-1\right),
\end{equation}
where $r$ is the radial coordinate and $M$ is the BH mass. The potential reads (see e.g. \cite{Birrell:1982ix})
\begin{equation}\label{SchwarzschildPotential}
	V(r)=\left(1-\frac{2M}{r}\right)\left(\frac{2M}{r^{3}}+
         \frac{\ell(\ell+1)}{r^{2}}\right),
\end{equation}
where $\ell$ is the angular momentum and $r$ is understood to be an implicit function of $r_*$.

Since we are only interested in high energy behavior, we exploit the complete analogy with quantum mechanics and use the Born approximation in which the reflection coefficient reads
\begin{equation}
   R(p)=\frac{1}{2i \, p}\int_{-\infty}^{\infty}V(r) \, e^{-2i \, p \, r_*}dr_*.
\end{equation}
For the potential \eqref{SchwarzschildPotential}, this has a closed form expression,
\begin{equation}
   R(p)=  -\Gamma(-4iMp)\left[2iMp \, \Gamma(4iMp,4iMp)+(4iMp)^{4iMp}e^{-4iMp} \left(\frac{1}{2} +\ell(\ell+1)\right)\right],
\end{equation}
where $\Gamma(a,z)$ is the incomplete gamma function. At high energies this behaves as
\begin{equation} \label{HighEnergySchwarzRefCoeff}
    R(p) \sim -\frac{i \pi}{2} e^{-\frac{\beta}{2} p},
\end{equation}
when written in terms of the temperature, $\beta = 8\pi M$. We see that, indeed, this obeys \eqref{l}.% and that the angular momentum enters only as a sub-leading contribution.

We note that an expression for the high frequency behavior of the reflection coefficient was already given in \cite{Sanchez:1976}. However, there, the solutions to the Klein-Gordon equation are given in terms of $r$ (and not $r_*$). Additionally, the asymptotic functions according to which the reflection coefficient is defined, were chosen to be $\exp{\left[\pm ip \left(r + 2M\log{(2pr)}  \right)\right]}$. This choice gives a reflection coefficient that differs from \eqref{HighEnergySchwarzRefCoeff}. However, upon making the switch to the tortoise coordinate \eqref{SchwarzTortoise} and the consequent redefinition of the reflection coefficient, one gets an expression which is in agreement with  \eqref{HighEnergySchwarzRefCoeff}.
\end{appendix}

\end{document}